# M Theory: Uncertainty and Unification

Joseph Polchinski

October 29, 2018

## 1  Introduction

We are in the middle of a series of important centennials: Wolfgang Pauli in 2000, Enrico Fermi and Werner Heisenberg in 2001, and Paul Dirac in 2002. This has presented an excellent opportunity to go back and review the scientific achievements of these men. Of course, the work that they did in the 20's, in their twenties, was their most important. But what I found more interesting was the work that they did afterwards. After they discovered quantum mechanics and established the basic framework of physics, they went on to try to understand the nuclear interaction, and quantum field theory, and the spectrum of particles and how all these things fit together. Some of these problems have since been solved, and it is interesting to compare their efforts with what we now know. Some of these problems we still struggle with, and it is even more interesting to compare the things that they tried with what we are trying today.

In many cases their point of view was surprisingly modern. Many of them tried to find a unified theory. Pauli, for one, was very attracted by Kaluza-Klein theory, the unification of gravity and electromagnetism in higher dimensions. Einstein, who was older of course, is well-known for his attempts at a unified theory, and Heisenberg is remembered for his attempts at a Worldformula.

Today many of us believe that there is a Worldformula. That is, there is a physical-mathematical structure that incorporates quantum mechanics, special relativity, general relativity, and the particles and their interactions, and which is beautiful and unique. We do not know the final form of this theory; we are like the quantum mechanicians in the early twenties, discovering the theory a piece at a time. In this talk I would like to present our current understanding of the Worldformula, M theory, and to structure the talk around some of the themes that were important in Heisenberg's work: a fundamental length, uncertainty, nonlinearity, and observables.

## 2  A Fundamental Length

Before getting to M theory, I want to say a few words about quantum field theory, one of the more-or-less solved problems. After quantum mechanics the next step was to incorporate special relativity. In principle this is straightforward and leads to quantum field theory. The problem was that the result had divergences, infinities. These arise because in quantum field theory the number of observables is infinite — for example, the values of the electric and magnetic fields at every point,

$$\vec{E}(\vec{x}) \ , \quad \vec{B}(\vec{x}) \ . \tag{1}$$

The discoverers of quantum mechanics thought very hard about this problem and tried many solutions. There are two broad classes of solution. One is that quantum field theory breaks down at some fundamental distance that I will call $l_0$. The other is this idea of renormalization, that the infinities do not appear in observables, they cancel and leave finite results.

According to most textbooks, the second idea won out, that it is through renormalization that quantum field theory makes sense. Many of the pioneers of quantum mechanics found this unattractive, and so it is worth emphasizing that our modern point of view is really a combination of these two approaches, and is actually closer to the first [1]. That is, the quantum field theories that we deal with are not valid down to arbitrarily short distance. Mathematically some of them (the asymptotically free ones) might make sense to arbitrarily short distance, but as a point of physics we don't expect them to be valid this far. At successively shorter scales one expects to encounter new quantum field theories, and ultimately no quantum field theory at all. The technical content of renormalization theory remains, but it has a new and much more physical interpretation: that the physics we see at long distances is largely independent of what is happening at very short distances, so we can calculate without knowing everything. I assume that many of the pioneers of renormalization thought in these terms, but this did not make it into the textbooks, which for decades presented our fundamental understanding of quantum field theory as

$$\infty - \infty = \text{physics} \ . \tag{2}$$

No!

So if there is a fundamental length scale, what is it? Heisenberg's idea was based on the weak interaction. The Fermi coupling $G_{\rm F}$, setting $\hbar = c = 1$, has units of length-squared. This length is about $10^{-15}$ cm, and Heisenberg identified this with the fundamental length. The reason is that at shorter distances $l$ the effective dimensionless coupling $l_0^2/l^2$ becomes large, and in Heisenberg's words physics becomes 'turbulent.'

Of course we now know that at the weak length scale, before turbulence can set in we just run into a new quantum field theory, Yang-Mills theory. But there is another constant of nature with units of length-squared, Newton's constant $G_{\rm N}$, where $l_0$ would be the Planck length $10^{-32}$ cm. Here we really do believe that there is a fundamental and final length scale, because when gravity becomes strong it is spacetime itself that becomes turbulent, and the notion of distance ceases to make sense. Whereas the weak interaction describes particles in a fixed spacetime, gravity describes spacetime itself, and so it is at here that Heisenberg's turbulence argument implies a fundamental length. As far as I know, Heisenberg never thought directly about quantum gravity, because he was focused on the microscopic world, but we have learned that in order to make progress we have to think about everything.

It is interesting to note that in the recent idea of large extra dimensions, the Fermi constant really does set the fundamental length scale. Things are more complicated because there is another length in the problem, the size $R$ of the extra dimensions. One then has

$$\begin{aligned} G_{\rm F} &= l_0^2 \,, \\ G_{\rm N} &= l_0^{2+n} R^{-n} \,, \end{aligned} \quad (3)$$

where $n$ is the number of large dimensions. I won't expand on this further, but it is curious that Heisenberg may have had the right length scale after all [2].

What happens at the fundamental scale $l_0$? In quantum field theory the interactions take place at spacetime points. When there is a fundamental length scale then the interactions must be spread out in some way, and this is not easy to do. It is not easy because there is a symmetry between space and time, special relativity, so if there is a spreading in space there is a spreading in time as well. Then there it the danger of losing causality and unitarity, so that physics does not make sense. In fact, in the case of quantum gravity, of everything that has been tried only one idea has worked, which is to replace the points with tiny loops, strings. And, strange as this is, string theory turns out to incorporate, and extend, many of the other unifying principles that have been tried and seem promising — supersymmetry, grand unification, and Kaluza-Klein theory.

I should mention that we often call the theory that we are working on string theory, because it has largely grown out of string theory, but it it has now grown into a larger structure. Thus we often call it M theory, a deliberately mysterious name for a theory whose final form we do not know.

## 3  Uncertainty

I am not going to describe string theory directly — this has been done in many other places — but I am going to approach it in a way that may be closer to Heisenberg's thinking. The idea of a fundamental length to which Heisenberg was so attached sounds like an uncertainty principle, but one that involves position alone and not momentum:

$$\delta x > l_0 \,. \quad (4)$$

This suggests that the fundamental length arises in the same way as the position-momentum uncertainty, that is that the coordinates do not commute with one another,

$$[x^\mu, x^\nu] \neq 0 \,. \quad (5)$$

One would certainly guess that Heisenberg would have tried this.[1]

---

[1] Jurg Fröhlich and other members of the audience confirmed this after the talk.

There are actually several ways to introduce such noncommuting coordinates. An obvious thing is to put some constant matrix on the right-hand side

$$[x^\mu, x^\nu] = \theta^{\mu\nu} ,\qquad(6)$$

so that spacetime becomes like a quantum mechanical phase space. The obvious problem is that the right-hand side is an antisymmetric tensor, so this cannot be Lorentz invariant; this is undoubtedly the main obstacle that inhibited the exploration of this direction. Nevertheless it is an interesting idea, which can be incorporated into quantum field theory and modifies the short-distance structure in puzzling ways (though it does not remove the short distance divergences) [3]. This kind of noncommutativity does appear in string theory, where the matrix $\theta^{\mu\nu}$ is the value of some spacetime field, but it only applies to the coordinates of open, not closed, strings. It is not clear what the role of this noncommutativity is, or how fundamental it is, since you can turn it off by setting the tensor field to zero. I should note though that Witten's open string field theory is in a sense an enlargement of this idea.

I would like to talk about another way to introduce noncommutativity of coordinates. Consider a nonrelativistic system of $N$ particles. Its configuration space is defined by the $N$ sets of coordinates

$$x_a^i ,\quad a = 1, \ldots, N ,\qquad(7)$$

where $i$ labels the coordinate axes and $a$ labels the different particles. Now let us make a different guess as to how to make these noncommutative. Let us double the lower, particle, index to make these into matrices,

$$x_{ab}^i ,\quad a,b = 1, \ldots, N .\qquad(8)$$

This is a bit strange, but it is not so far from the spirit of how Heisenberg guessed at matrix mechanics, so let us try it and see where it leads. Now in general the different spatial coordinates do not commute:

$$\sum_b (x_{ab}^i x_{bc}^j - x_{ab}^j x_{bc}^i) \neq 0 ,\qquad(9)$$

so this is another way to introduce noncommutativity into the coordinates. In a sense it makes the particle identities uncertain as well, because we can now change the basis for the matrices.

We want physics at low energy to have its familiar form, while the new noncommutativity becomes important at high energy. We can arrange this by adding a certain potential energy term to the Hamiltonian. Here is the Hamiltonian, written in matrix notation:

$$H_0 = \frac{1}{l_0} \sum_i \mathrm{Tr}(\dot{\mathbf{x}}^i \dot{\mathbf{x}}^i) + \frac{1}{l_0^5} \sum_{i,j} \mathrm{Tr}([\mathbf{x}^i, \mathbf{x}^j]^\dagger [\mathbf{x}^i, \mathbf{x}^j]) .\qquad(10)$$

The first term is an ordinary kinetic term for every component of every matrix. The second term is the sum of the squares of every commutator (9), so that at low energy these commutators must vanish and we recover the ordinary commuting positions; at high energy the noncommutativity appears. Then this simple Hamiltonian has the desired property.

Actually, this doesn't quite work yet, because the quantum corrections spoil the structure, in that they produce a nonzero energy even when the coordinates commute. But we know of a general way in physics to cancel quantum corrections, and that is to introduce supersymmetry. One introduces in addition to the real number coordinates $x^i_{ab}$ some fermionic coordinates $\psi^i_{ab}$; essentially this means that the particles can have various spins. Adding an appropriate coupling of the real and fermionic coordinates to the Hamiltonian,

$$H = H_0 + \frac{1}{l_0^2} \sum_i \text{Tr}(\psi \gamma^i [\mathbf{x}^i, \psi]) \ , \tag{11}$$

makes the theory supersymmetric and cancels the unwanted quantum corrections. The theory then behaves as desired, commutative at low energy and noncommutative at high energy.

Once we have added supersymmetry it is natural to consider the largest possible supersymmetry algebra. It happens that the largest possible algebra has 16 supersymmetry charges. But once we take this step, we begin to encounter a nice convergence of ideas: the funny commutator potential term, which we added in order to get back ordinary physics at low energies, is in fact the *unique* potential allowed when there are 16 supersymmetries! This is a sign that we are on the right track — we are getting more out than we put in.[2]

In fact, things are even better. If we look now at the low energy physics of the commuting coordinates, the noncommuting parts of the coordinate matrices give virtual effects. One can calculate this, and one finds that the net effect of the virtual degrees of freedom is precisely to give a *gravitational* interaction (or supergravitational, to be precise) between the particles. Gravity is not put in from the start, it is a derived effect of the noncommutativity!

Could it be that eq. (11), and not string theory, is the Worldformula? Yes, and no. Eq. (11) very likely is the Worldformula, but it is not an alternative to string theory, it *is* string theory. To be precise, this is the Banks-Fischler-Shenker-Susskind matrix theory, describing M theory in eleven asymptotically flat dimensions with one of the null directions periodic [4]. So this is a formula for a world, but not for our world. It is a complete description of one sector of the Hilbert space of M theory, but one that still has a lot of physics — gravitons, black holes, strings, and branes are all described by this simple matrix Hamiltonian. We live in a much less symmetric state, where seven of the dimensions are curved and compact, and on top of this the geometry of our spacetime is changing in time. We do not yet know the correct form of matrix theory or M theory in our much less symmetric state, it is undoubtedly much more complicated.

---

[2] I should note that the symmetry also fixes the number of spatial dimensions; the number is nine, not three, but again we have known since Kaluza and Klein that the existence of extra space dimensions is a powerful unifying principle.

# 4  Nonlinearity

So how do we see the strings and branes in the Hamiltonian (11)? Essentially, the particles can link up, due to their noncommutative nature, into loops and higher-dimensional structures. It is essential here that the Hamiltonian is nonlinear. This was an important part of Heisenberg's thinking also, that we could start from a simple Hamiltonian and build up complicated physics via nonlinearities. QED is a nice textbook example of a weakly coupled field theory, where the nonlinearities can be treated perturbatively, but the most interesting phenomena in physics, like quark confinement, dynamical symmetry breaking, and black holes, arise due to strong nonlinearities.

One of the important things that we have learned in the past few years is that nonlinear theories do not have to be ugly and chaotic. For the particular Hamiltonians that arise in string theory and M theory, it happens in many cases that just when the nonlinear effects become very large, and you would expect that the physics becomes very 'turbulent,' there is a new set of variables in terms of which the physics becomes approximately linear. This is called a 'duality,' and it is a remarkable phenomenon that has enabled us to make great progress in understanding string/M theory. For example, the matrix Hamiltonian (11) can be recast in terms of string variables, and the theory takes the familiar form of a sum over string world-histories; this string description becomes weakly coupled (linear) when exactly one of the coordinates $x^i$ is made periodic.

One way to summarize our understanding of string theory is through a sort of phase diagram, shown in Fig. 1 [5]. In various limits, which are the corners of the diagram, the physics linearizes. Five of these points correspond to one or the other of the string theories, and the sixth is the eleven-dimensional theory that I have been discussing. Up until a few years ago, all we understood was the five stringy points and their neighborhoods, but now we are able to map out the whole diagram. What we used to think of as different theories are just different phases in a single theory.

If this were the phase diagram of water, say, then the parameters would be the pressure and temperature. Here, the parameters are the shapes and sizes of the compact dimensions. M theory has gravity, so spacetime is dynamical. We are most interested in spacetimes like ours, which has four large spacetime dimensions and the rest small and compact. Even if we cannot see directly those compact dimensions, the important principle is that the physics that we do see depends on their geometry and topology. So it is this geometry that is varying as we move around the diagram, and there are certain limits of the geometry in which the physics becomes linear in some set of variables. By the way, this diagram is greatly oversimplified, in that there are many parameters and many more pieces of the diagram which join each other across phase transitions. When a four-dimensional physicists sees a phase transition, a qualitative change in the physics, what is usually happening from the higher-dimensional point of view is a change of the topology of space.

In the middle of the diagram, away from the linear limits, we do not know how to calculate, but what is worse is that we do not know even in principle what calculation to do, we do not know what the theory is. We do not know the full Hamiltonian, and we do not even know what variables it should be written in terms of. The variables are almost certainly not strings, one-dimensional objects. We have always suspected this, but with the understanding of duality it is clear that the string variables are useful only to expand around special limits of the phase diagram, and in other phases and other descriptions the variables are very different.

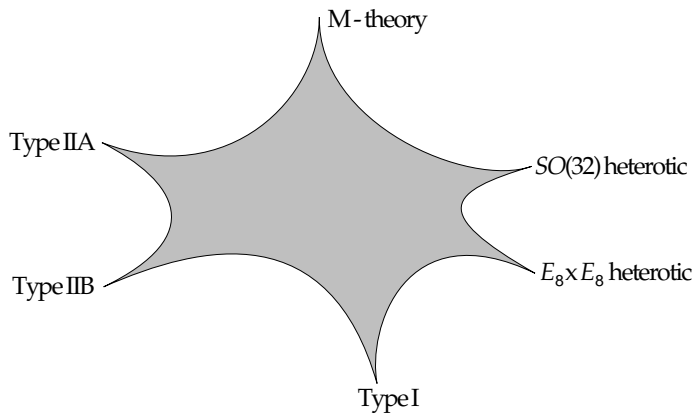

Figure 1: A piece of the phase diagram of M theory

## 5 Observables

For all we understand string/M theory, we still do not know its central defining principle, the analog of the uncertainty principle in quantum mechanics and the equivalence principle in general relativity. What we need is for one of the young people in the audience to do what Heisenberg did, to go off to Heligoland for a few weeks and figure it out. Before you go, I would like to try to play the role of Bohr, and give you a few things to think about.

First, the key step may be to identify what are the physical observables, and what cannot be observed. For example, the equivalence principle tells us that we cannot measure absolute velocity or absolute acceleration. The uncertainty principle tells us that we cannot measure position and velocity to arbitrary accuracy.

In string/M theory, the issue of observables has been around for a while. The obvious observable in string theory has always been the S matrix, the amplitude to go from some configuration of strings (or strings and branes) in the infinite past to some other configuration in the infinite future. This correctly incorporates the principle that we can only make measurements with physical objects. For example, we cannot talk about some local operator at a point without a prescription for measuring it in a scattering experiment. On the other hand, the S matrix does not correspond to our experience of time in an ongoing way. It is even more a problem in cosmology, where the universe may not have an infinite past and future.

It is worth noting at this point that Heisenberg is in a rather direct sense the great-grandfather of string theory:

$$\text{Heisenberg} \to \text{Chew} \to \text{Veneziano} \to \text{strings} . \qquad (12)$$

The strong interaction was a difficult problem for a very long time, and one of the ways that Heisenberg tried to approach it, in the 40's, was via the same route that he understood quantum mechanics: identifying the physical observables. So he invented the S matrix for just this purpose, and he further proposed that it would be determined entirely by physical consistency, unitarity and analyticity.

Heisenberg dropped this idea a few years later, in favor of a more dynamical approach. But the strong interaction remained unsolved twenty years later, and so Chew and others returned to the idea that we should consider only the S matrix and its consistency conditions. For the strong interaction this was not correct, it is a local field theory, but it led Veneziano to make an inspired guess and write down a simple solution to the consistency conditions. His model was interpreted a few years later as describing a theory of strings, and that led in turn to strings as a theory of gravity and everything else.[3]

So the issue of observables has been central to the history of string theory, and it is probably also a key to its future.

## 6 On to Heligoland

We do have an idea of what the central principle is, and we call it the holographic principle. We do not have a precise formulation of this, but the rough statement is that if we have a system in some region, the states of the system can be characterized by degrees of freedom living on the surface of that region [6]. This is completely contrary to our experience and to quantum field theory, where the degrees of freedom would live at points in the interior of the region. But there are strong arguments that this must be true in a theory of quantum gravity, and it is much less local than one would have with just a minimum length. It means that the thing that we must give up in our next revolution is the underlying locality of physics.

This principle is suggested by black hole quantum mechanics, where the entropy is proportional to the surface area. It has a precise realization in recent dualities in string theory, the AdS/CFT duality and generalizations, where the states of string theory in the bulk of the anti-de Sitter spacetime are isomorphic to the states of gauge fields on the boundary. However, anti-de Sitter spacetime is very special, and the realization of the holographic principle in more general settings is not known.

---

[3]Helmut Rechenberg, curator of the Werner Heisenberg archive, has informed me that the chain (12) is even more direct than I had guessed. As early as 1954, Heisenberg wrote in a letter that in Urbana he had met 'a particularly nice younger physicist with the name Chew.' Also, the famous Regge pole paper was also written at Heisenberg's Munich institute.

Many of the open puzzles in string theory seem to center on cosmology:

- Why is the cosmological constant so small, and why then is it not exactly zero?

- What are the observables in a cosmological situation, and how does one formulate the holographic principle, especially if the spatial geometry is closed?

- How are cosmological singularities resolved? This is a problem that has been solved in string theory for many static singularities.

- How do we find a unified theory of the dynamical laws and the initial conditions?

I have presented this as a purely theoretical discussion; unfortunately experiment still gives little guidance as to what lies beyond the Standard Model, and what is the theory of quantum gravity. Notice, however, that the apparent observation of a positive cosmological constant has very strongly affected the thinking of string theorists. In particular, it very much complicates the formulation of the holographic principle. So even a small amount of data can have a large impact. Let me therefore echo Michael Peskin's message about the importance of building TESLA.

Finally, let me wish the young people in the audience: have a good trip to Heligoland, and call when you get back!

## Acknowledgements

I would like to thank Jurg Fröhlich and Helmut Rechenberg for their comments. This work was supported by National Science Foundation grants PHY99-07949 and PHY00-98395.

Given the wide span of this talk, I list below only a few review articles for those who wish to pursue some subjects further.